\documentclass[aps,prl,twocolumn,amsmath,amssymb,superscriptaddress,nobibnotes,showpacs,floatfix]{revtex4-1}
\usepackage{graphicx}
\pdfoutput=1
\usepackage{ulem}
\usepackage{amsmath}
\usepackage{amssymb}
\usepackage{amsthm}
\usepackage{tabularx}
\usepackage{multirow}

\usepackage[usenames, dvipsnames]{color} 

\usepackage{hyperref}
\hypersetup{
  colorlinks = true,    
  allcolors = black
}

\usepackage[usenames, dvipsnames]{color} 

\usepackage{changes}
\definechangesauthor[name=Markus Aichhorn, color=red]{ma}

\usepackage{hyperref}
\hypersetup{
  colorlinks = true,    
  allcolors = black
}

\begin{document}
\newcommand{\ma}[1]{\textcolor{red}{\textbf{[NOTE: #1]}}}
\title{The Mott transition in the 5d$^1$ compound Ba$_2$NaOsO$_6:$ a DFT+DMFT study with PAW  spinor projectors}

\author{Dario Fiore Mosca}
\affiliation{Centre de Physique Th\'eorique, Ecole Polytechnique, CNRS, Institut Polytechnique de Paris,
  91128 Palaiseau Cedex, France}
\affiliation{Coll\`ege de France, 11 place Marcelin Berthelot, 75005 Paris, France}

\author{Hermann Schnait}
\affiliation{Institute of Theoretical and Computational Physics, Graz University of Technology, NAWI Graz, 8010 Graz, Austria}

\author{Lorenzo Celiberti}
\affiliation{University of Vienna, Faculty of Physics and Center for Computational Materials Science, Vienna, Austria}

\author{Markus Aichhorn}
\affiliation{Institute of Theoretical and Computational Physics, Graz University of Technology, NAWI Graz, 8010 Graz, Austria}

\author{Cesare Franchini}
\affiliation{University of Vienna, Faculty of Physics and Center for Computational Materials Science, Vienna, Austria}
\affiliation{Department of Physics and Astronomy "Augusto Righi", Alma Mater Studiorum - Universit\`a di Bologna, Bologna, 40127 Italy}

\date[Dated: ]{\today}

%

\begin{abstract}
Spin-orbit coupling has been reported to be responsible for the insulating nature of the 5d$^1$ osmate double perovskite Ba$_2$NaOsO$_6$ (BNOO). However, whether spin-orbit coupling indeed drives the metal-to-insulator transition (MIT) in this compound is an open question. In this work we investigate the impact of relativistic effects on the electronic properties of BNOO via density functional theory plus dynamical mean-field theory calculations in the paramagnetic regime, where the insulating phase is experimentally observed. The correlated subspace is modeled with  spinor projectors of the projector augumented wave method (PAW) employed in the Vienna Ab Initio Simulation Package (VASP), suitably interfaced with the TRIQS package. The inclusion of PAW  spinor projectors in TRIQS enables the treatment of spin-orbit coupling effects fully ab-initio within the dynamical mean-field theory framework. In the present work, we show that spin-orbit coupling, although assisting the MIT in BNOO, is not the main driving force for its gapped spectra, placing this material in the Mott insulator regime. Relativistic effects primarily impact the correlated states' character, excitations, and magnetic ground-state properties. 
\end{abstract}

\maketitle

\section{I. Introduction}

Mott insulators are a prominent class of materials, predominantly found in 3d transition metal oxides (TMOs), where the spatially localized nature of the d orbitals enhances correlation effects~\cite{imada1998metal}. The Mott metal-insulator-transition (MIT) involves the competition between a strong electron-electron repulsion ($U$) and the kinetics of electrons, represented by the bandwidth ($W$)~\cite{mott1968metal}. When interactions are strong enough, i.e. $U/W$ large, charge carriers can localize in systems where conventional band theories would predict metallic states~\cite{imada1998metal}. 
Conversely, 4d and 5d TMOs exhibit more delocalized orbitals, larger bandwidths, and an overall lower value of $U$. At first sight, this results in a small $U/W$ ratio, and the materials should show metallic properties, as observed in  SrRuO$_3$~\cite{RevModPhys.84.253}. 
In contrast to the expected behavior,  5d TMOs like Sr$_2$IrO$_4$  display insulating character if, as demonstrated,  the conditions of unfilled shells and strong spin-orbit coupling (SOC) effects in a cubic crystal field are met~\cite{PhysRevLett.101.076402,sohn2014orbital}. This phenomenon is now known as the relativistic-Mott or Dirac-Mott insulating phase. It originates from the modification of the atomic levels in the presence of SOC, which causes a different filling of the new spin-orbital states, enhancing correlation effects~\cite{PhysRevLett.101.076402}.

Since the discovery of the SOC-driven MIT, other compounds have been reported to display similar properties,
including BNOO~\cite{Gangopadhyay2015,Gangopadhyay2016}. BNOO is a double perovskite with $Fm\bar{3}m$ space group and geometrically frustrated lattice. The single electron of the Os$^{7+}$ ion is coupled via a strong SOC effect ($\lambda~\sim$~0.3 eV~\cite{mosca2021interplay}) to the unquenched low-lying t$_{2g}$ multiplet with effective angular momentum $\Tilde{l}=1$. Consequently, the one-electron levels, now described by the effective total angular momentum operator, split into an excited doublet $J_\text{eff}=1/2$ and a ground-state quartet with $J_\text{eff}=3/2$. 
Strong electron-electron repulsion ($U\sim 3.3$\,eV~\cite{Erickson2007}) and the Jahn-Teller (JT) active ground-state multiplet provide the remaining ingredients for the observation of novel physics~\cite{Chen2010, takayama, lu2017magnetism, mosca2021interplay, celiberti2023spinorbital}. The magnetic ground state of BNOO is of type-I canted antiferromagnetic order originating from complex multipolar interactions coupled with local JT distortions~\cite{Chen2010,mosca2021interplay,lu2017magnetism}. 

\begin{figure}[t]
\includegraphics[width=1.0\columnwidth]{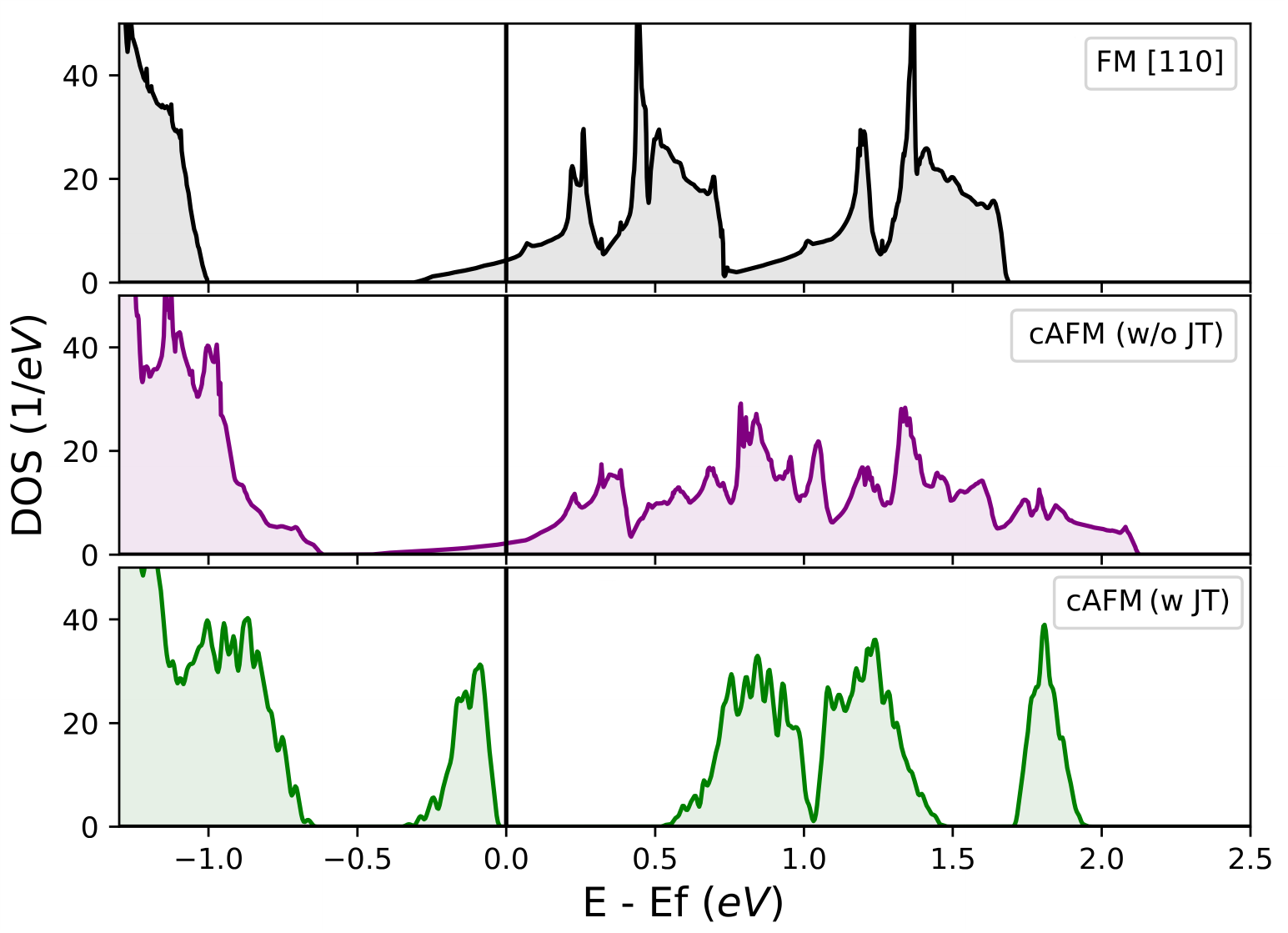}
\caption{The DOS of BNOO in an 8 f.u. supercell without SOC is shown for the cases of fully ferromagnetic solution with magnetic moments along [110]  \textcolor{black}{ as in Ref.~\cite{Gangopadhyay2015} (top panel), for the cAFM phase without JT distortions (middle  panel), and for the JT-distorted cAFM structure (bottom panel). The cAFM magnetic configuration is taken from Ref.~\cite{mosca2021interplay}. For the computational details and the JT-distorted structure see the Supplementary materials II}.
}
\label{fig:0} 
\end{figure}

Moving to the electronic properties, both DC resistivity and infrared reflectivity measurements indicated that BNOO is an insulator at room temperature~\cite{Erickson2007}. Earlier results from density functional theory (DFT) calculations on undistorted unit cells claimed for the Dirac-Mott type of transition, showing that a gap can be exclusively opened when including relativistic corrections~\cite{Gangopadhyay2015, Gangopadhyay2016, Xiang2007}. 
The authors show, however, that already at the DFT+U+SOC level, the insulating phase strongly depends on the underlying magnetic configuration~\cite{Gangopadhyay2015, Gangopadhyay2016}. 
Later studies have further highlighted the complex interplay of JT effect, electronic correlation and orbital ordering in the onset of the zero-temperature gapped spectra, proving how these degrees of freedom are highly intermingled in this system~\cite{PhysRevB.100.245141,PhysRevB.97.224103}.
Building upon these previous studies, \textcolor{black}{we computed the Density of States (DOS) in DFT on supercell structures deliberately excluding SOC. This approach revealed that the canted antiferromagnetic (cAFM) ground state can exhibit either metallic or insulating phase, depending on whether Jahn-Teller (JT) distortions are included or not (see Fig.~\ref{fig:0}). We note that the JT solution has been derived without incorporating SOC.
This was done intentionally to illustrate that a gap in BNOO can be opened without invoking on relativistic effects. Additional details on the JT phase can be found in Supplementary Information II. } 
Moreover, the onset of the magnetic ordering and local symmetry breaking occurs at $T_N \approx 6.8$\,K and T$~\approx 10$\,K~\cite{Erickson2007, lu2017magnetism, PhysRevB.100.041108, LIU2018863} respectively, i.e., far below the experimental observation of the insulating phase at 300 K to justify their impact on the gapped spectra in the high-temperature paramagnetic phase.
These ambiguities call for a clarification of the main force driving the observed insulating state in BNOO.

In this work, we are interested in understanding the insulating nature of BNOO and deciphering SOC's role in its MIT. To accomplish this, we utilize a combination of DFT and dynamical mean-field theory (DFT+DMFT) calculations. Specifically, we focus on the room-temperature paramagnetic regime, which is well above the magnetic phase transition and is characterized by a perfectly cubic phase without JT distortions. To account for SOC effects from VASP bandstructure calculations~\cite{Kresse_1994,PhysRevB.54.11169, PhysRevB.47.558}, we implemented an extension of the plovasp converter included in the TRIQS/DFTTools~\cite{Parcollet_2015_PAPER,Aichhorn2016,triqs_wien2k_interface} package that encompasses  spinor PAW projectors. We tested its accuracy by comparing our results with the equivalent linearized augumented plane wave (LAPW) projectors of Wien2k, showing that VASP  spinor projectors allow treating, fully ab initio, SOC effect in DFT+DMFT calculations. Furthermore, this implementation also provides the opportunity to explore magnetic non-collinear phases within DFT+DMFT.

Moving back to BNOO,  we show that SOC is not indispensable for opening the gap, as its insulating phase is  equivalently reproduced without relativistic corrections for both values of $U$ extracted from experiments and computed with constrained Random Phase Approximation (cRPA)~\cite{PhysRevB.70.195104}. The Mott insulating nature is thus predominant in this material. But while SOC does not play a leading role in gap formation, it significantly influences the nature of its atomic states, now described by $J_\text{eff}$ states, resulting in the observation of the exotic magnetic ground state~\cite{mosca2021interplay}, as well as in the appearance of an additional feature in the upper Hubbard band (UHB).

The paper is structured as follows: Section \href{Sec:2a}{II A} summarizes the  spinor implementation of SOC in VASP, the  spinor extension of the PAW projectors, and DMFT equations are in Sec. \href{Sec:2b}{II B} and \href{Sec:2c}{II C} respectively. Section \href{Sec:3}{III} describes the computational procedure for the DFT+DMFT calculations. In Sec. \href{Sec:4a}{IV A} we compare the treatment of the correlated subspace with Wien2k~\cite{Blaha_2020_PAPER}; in Sec. \href{Sec:4b}{IV B} we present the electronic properties of BNOO and discuss its MIT. Section \href{Sec:5}{V} offers some final remarks.

\section{II. Methods}

\subsection{A.  Spinor PAW and SOC}
\label{Sec:2a}

Spin-orbit coupling in VASP~\cite{PhysRevB.54.11169,PhysRevB.47.558} is treated within the  spinor PAW framework. Here we consider the corresponding extension of the projector augmented wave method~\cite{Kresse_1994,Hobbs}. In its original derivation~\cite{doi:10.1063/1.340744,Kubler_1988}, it describes Kohn-Sham states in terms of spinor wavefunctions such that the general density matrix reads
\begin{equation}
    n^{\alpha\beta}(\mathbf{r}) = \big[ n_\text{Tr} (\mathbf{r}) \delta_{\alpha\beta} + \vec{m}(\mathbf{r}) \cdot \Vec{\sigma}^{\alpha\beta} \big] /2 \ .
    \label{eq:1}
\end{equation}
where $n_\text{Tr}(\mathbf{r})$ is the electron density defined as  
\begin{equation}
    n_\text{Tr}(\mathbf{r}) =  \text{Tr}[n^{\alpha \beta} (\mathbf{r}) ]  = \sum_{\alpha} n^{\alpha \alpha} (\mathbf{r}) 
\end{equation}
and  $\Vec{m}(\mathbf{r})$ is the magnetization density 
\begin{equation}
    \Vec{m}(\mathbf{r}) = \sum_{\alpha \beta} n^{\alpha \beta} (\mathbf{r}) \cdot \Vec{\sigma}^{\alpha\beta}
\end{equation}
with $\vec{\sigma} = (\sigma_x, \sigma_y, \sigma_z)$ the $2\times2$ Pauli spin matrices~\cite{Hobbs}. The indices $\alpha, \beta$  describe the spinor components from now on.   
In VASP, with the PAW formalism, the one-electron (spinor) wavefunctions $|\Psi^{\alpha}\rangle$ can be decomposed exactly as~\cite{Karolak2011}
\begin{align}
    |\Psi^{\alpha} \rangle &= \mathcal{T} |\Tilde{\Psi}^{\alpha} \rangle =  \Big( 1 + \sum_i \tau_{i} \Big) |\Tilde{\Psi}^{\alpha} \rangle \\
    & = |\Tilde{\Psi}^{\alpha} \rangle + \sum_i \Big(|\phi_i\rangle - |\Tilde{\phi}_i \rangle \Big) \langle \Tilde{p}_i | \Tilde{\Psi}^{\alpha} \rangle\, ,
\end{align}
where the pseudo orbitals (PS) $ | \Tilde{\Psi}^{\alpha} \rangle$ are the variational quantities of the Kohn-Sham equations,  $|\phi_i \rangle$ are the all-electron (AE) partial waves for the non-magnetic ion, $|\Tilde{\phi}_i \rangle$ are equivalent to the AE partial waves outside a core radius and continuously match the AE waves inside the core.  Lastly, the projector functions  $\Tilde{p}_i$ are chosen  such that $\langle \Tilde{p}_i | \Tilde{\phi}_j \rangle~=~\delta_{ij}$. 

The corresponding Kohn-Sham equations are obtained through the application of the variational principle to the total energy functional, whose result can be written in a compact way as
\begin{equation}
    \sum_\beta H_{KS}^{\alpha\beta} |\Tilde{\Psi}_{n}^{\beta} \rangle = \varepsilon_{n} S^{\alpha \alpha} |\Tilde{\Psi}_{n}^{\alpha} \rangle
\label{eq:6}
\end{equation}
where $n = \{\mathbf{R},l, m_l, ...\}$  is an index for the set of one-electron quantum numbers and $S^{\alpha \alpha}$ is the overlap operator $\langle \Tilde{\Psi}^{\alpha}_{m} | S^{\alpha\alpha} |\Tilde{\Psi}^{\alpha}_{n} \rangle = \delta_{mn} $~\cite{Hobbs}. Eq.~\eqref{eq:6} is the generalized Kohn-Sham equation for the pseudo-wave functions~\cite{Hobbs}.

Moving to the SOC effect, its expression is found to be, for an electron with rest mass $m_e$ in presence of a potential $V(r)$~\cite{Steiner2016, lenthe1993relativistic},
\begin{equation}
    H^{\alpha \beta}_{SOC} = \frac{\hbar^{2}}{(2m_{e}c)^2} \frac{K(r)}{r}\frac{dV(r)}{dr} \ \Vec{\sigma}^{\alpha \beta} \cdot \Vec{L}
\end{equation}
where $\Vec{L} = \Vec{r} \times \Vec{p}$ is the orbital angular momentum operator, $c$ the speed of light, $\hbar$  the reduced Planck's constant, and  
\begin{equation}
    K(r) = \Bigg( 1 - \frac{V(r)}{2mc^2} \Bigg)^{-2} \ .
\end{equation}
The important point is that the action of $H_{SOC}$ on the one-electron orbitals is restricted to the pseudo-waves, due to the semi-locality of the operator itself, whose action is negligible outside the PAW spheres, such that~\cite{Steiner2016}
\begin{equation}
    \textcolor{black}{| \Tilde{\Psi}^{\alpha}_n \rangle = \sum_{ \beta} \Tilde{H}^{\alpha\beta}_{SOC} | \Tilde{\Psi}^{\beta}_{n} \rangle \ },
\end{equation}
 and 
\begin{equation}
    \Tilde{H}_{SOC} = \sum_{ij} |\Tilde{p}_i \rangle \langle \phi_i | H_{SOC} | \phi_j \rangle \langle \Tilde{p}_j | \ .
\end{equation}

\subsection{B.  Spinor projected localized orbitals}
\label{Sec:2b}
In DFT+DMFT we express the many body quantities into a local-orbital Wannier-like basis set~\cite{Schuler2018}. 
It has been demonstrated that projection operators provide a reliable procedure for performing such unitary transformation. One technique that utilizes projection operators is referred to as projected localized orbitals (PLO)~\cite{Schuler2018,Karolak2011,Amadon2008}.
Following the derivation of Ref.~\cite{Schuler2018}, we can define spinor PLOs as an orthonormal basis set $|\chi^{\alpha}_{L} \rangle$ that spans the correlated subspace $\mathcal{C}$ at each site, where $L = (l,m_l)$ represents the set of local quantum numbers. With $\langle \chi^{\alpha}_{L} | \chi^{\beta}_{L'} \rangle = \delta_{\alpha \beta} \delta_{LL'}$ any operator $\hat{A}$ can be projected onto the correlated subspace $\mathcal{C}$ as 
$\hat{A}^{\alpha \beta}_{imp} = \hat{P}^{\alpha} \hat{A} \hat{P}^{\beta}$ with the projection operator~\cite{Schuler2018}
\begin{equation}
    \hat{P}^{\alpha} = \sum_{L} |\chi^{\alpha}_{L} \rangle \langle \chi^{\alpha}_{L} |\, ,
\end{equation}
and a vector of the Hilbert space $|\Psi \rangle$ can be decomposed as $ \hat{P}^{\alpha} | \Psi \rangle = \sum_{ L} |\chi^{\alpha}_{L} \rangle \langle \chi^{\alpha}_{L} | \Psi \rangle$.
By considering a complete basis set $\{ |\Psi_{\nu\mathbf{k}} \rangle \}$, the PLO functions allow to define the projector operator by expressing  
$ P^{\alpha}_{L,\nu} (\mathbf{k}) =  \langle \chi_{L}^\alpha | \Psi_{\nu\mathbf{k}} \rangle$. 
By extending the demonstration of ref.~\cite{Schuler2018,Karolak2011}, we can rewrite the spinor PLO projectors in the PAW formalism as 
\begin{equation}
    P^{\alpha}_{L,\nu} (\mathbf{k}) = \sum_i \langle \Tilde{\chi}_{L} | \phi_i \rangle \langle \Tilde{p}_i | \Tilde{\Psi}^{\alpha}_{\nu\mathbf{k}} \rangle \ ,
\end{equation}
where we have decomposed $| \chi^{\alpha}_L \rangle = | \Tilde{\chi}_L \rangle \otimes |\alpha \rangle$ into the product of an orbital-only and spinor-only components, while $i$ runs over the PAW channels. The spinor PLOs allow to rewrite the charge density of Eq.~\ref{eq:1} as: 
\begin{equation}
    \textcolor{black}{n^{\alpha \beta} (\mathbf{r}) = \sum_{n, \mathbf{k}} f_{n \mathbf{k}} \sum_{L} \langle \Tilde{\Psi}^{\alpha}_{\mathbf{k}} | \chi^{\alpha}_{L} \rangle \langle \chi^{\beta}_{L} | \Tilde{\Psi}^{\beta}_{\mathbf{k}} \rangle \ .} 
\end{equation}
where $f_{n \mathbf{k}}$ are the Kohn-Sham occupation numbers for band index $n$ and momentum $\mathbf{k}$. 
The VASP code has already implemented some projector schemes. Examples are the hydrogen-like functions or directly the all-electron partial waves. The overall performance varies, however, depending on the specific system/orbital. This freedom of choice finds its main drawback in its arbitrariness, as well as on the user-based decision on which projector to utilize. The problem has been addressed by Sch\"uler and coworkers~\cite{Schuler2018} by defining \textit{optimised} projectors via diagonalization of the all-electron one-center overlap matrix. This methodology is implemented in VASP and in the \textit{plovasp} converter of TRIQS/DFTTools. In this work we have extended it to account for  spinor projectors, readily available in VASP, and now as well in TRIQS/DFTTools~\cite{Aichhorn2016}.

\subsection{C. Non-collinear DMFT equations}
\label{Sec:2c}

The DFT+DMFT  fundamental equations are the ones relating the Kohn-Sham quantities, projected onto the correlated subspace, to the Anderson impurity problem~\cite{Amadon2012}. Within this framework, it is straightforward to extend the interacting Green's function to the  spinor case. In the following, to keep the notation light, we suppress the spinor indices $\alpha$, $\beta$, and assume implicitly that all sums include the spinor degree of freedom. The Greens function is then given by  
\begin{equation*}
    \hat{G}(\mathbf{k}, i\omega_{n})  = \Big[(i\omega_n + \mu) \hat{1} - \hat{H}_{KS}(\mathbf{k}) - \hat{\Sigma}_{KS} (\mathbf{k}, i\omega_n) \Big]^{-1} \ .
\end{equation*}
The self energy is obtained in complete analogy to Ref.~\cite{Schuler2018} by upfolding the local self energy. Since we do not assume the spin as a good quantum number, we treat the local problem as a $2 \mathcal{C} \times 2 \mathcal{C}$ matrix valued problem. The projection operators have matrix elements $P_{m\nu}$, as introduced above, where $\nu$ runs over all Kohn-Sham bands inside the projection window (taking the spinor degree of freedom into account). The indices $m$ and $m'$ run over the 2$\mathcal{C}$ local degrees of freedom, and we can write for the self energy
\begin{equation*}
\hat{\Sigma}_{KS} (\mathbf{k},i\omega_n)  = \sum_{m,m'} \Big(P_{\nu m}(\mathbf{k})\Big)^* \Sigma_{mm'} (i\omega_n) {P}_{m'\nu'} (\mathbf{k}) \ .
\end{equation*}
The TRIQS and TRIQS/DFTTools packages are written in a very flexible way, such that there is no specific requirements or restrictions on the number of orbitals in the correlated space, defining the dimension of the interacting Green's function and self energy, or on the number of Kohn-Sham eigenvalues in the projection window. Solving the Anderson impurity problem within the  spinor framework requires only a different initialization of the interacting Hamiltonian (see supplementary material). Its solution, i.e., the calculation of the interacting Greens functions and self energies will depend as usual on the choice of the impurity solver, since different impurity solvers will perform differently when dealing with a $2^{2 \mathcal{C} \times 2 \mathcal{C}}$  Hilbert space.

\section{III. Computational Details}
\label{Sec:3}

The DFT+DMFT calculation is performed in two steps. First, the DFT calculation gives the non-interacting Hamiltonian and the projectors, which are combined with an interaction Hamiltonian to serve as inputs in the DMFT self-consistent cycle. 

\textit{DFT setup}: The calculations were done with the primitive cell of the cubic-conventional fcc structure of Ref.~\cite{Stitzer2002} with lattice constant $a = 8.287$\,\AA. We  made use of the Perdew, Burke and Ernzerhof (PBE) treatment of exchange-correlation functional~\cite{perdew1996rationale}, and an energy cutoff of 600\,eV was applied together with an energy convergence factor of $10^{-7}$\,eV. This choice is necessary for a well-converged ($\sim 10^{-3}$) norm of the residuum of the wavefunctions. High accuracy in this quantity improves the quality of the projectors. The Brillouin zone was sampled with a k-mesh of $10 \times 10 \times 10$  while, for higher-precision calculations, a $14 \times 14 \times 14$  k-mesh was used. All magnetic moments were set to zero and all symmetries were switched off. The latter is a mandatory requirement for  spinor calculations, which were done also in absence of SOC to further test the correctness of the projectors. 

Regarding the many-body treatment of the correlated subspace, the optimal PAW  spinor PLOs were chosen within an energy window $\mathcal{W} = [-1, 8]$\,eV with respect to the Fermi level. In this way, we included e$_g$ states, whose mixing with t$_{2g}$ orbitals contributes to the total angular momentum eigenstates, as the e$_g$ and t$_{2g}$ orbitals are not anymore exact irreducible representation in presence of SOC~\cite{Schnait2022}.

\textit{DMFT setup}: The DMFT self-consistent cycle was performed with the TRIQS/DFTTools toolkit~\cite{Parcollet_2015_PAPER,Aichhorn2016}. The interactions were included in a Slater-type Hamiltonian with interaction values $U_S=3.0$\,eV and $J_S~=~0.48$\,eV for both relativistic and non-relativistic calculations, motivated by earlier experimental and theoretical works~\cite{Erickson2007,mosca2021interplay}. 
\textcolor{black}{When one projects out the $e_g$ subspace, this is equivalent  to work in the $t_{2g}$ subspace with a Kanamori Hamiltonian with values of $U=3.55$\,eV and $J=0.37$\,eV. Here we will make use of the Kanamori notation from now on. }
The calculations were performed at room temperature ($1/k_BT\approx 40$\,eV$^{-1}$) in the paramagnetic phase. 
\begin{figure}[b]
\includegraphics[width=0.95\columnwidth]{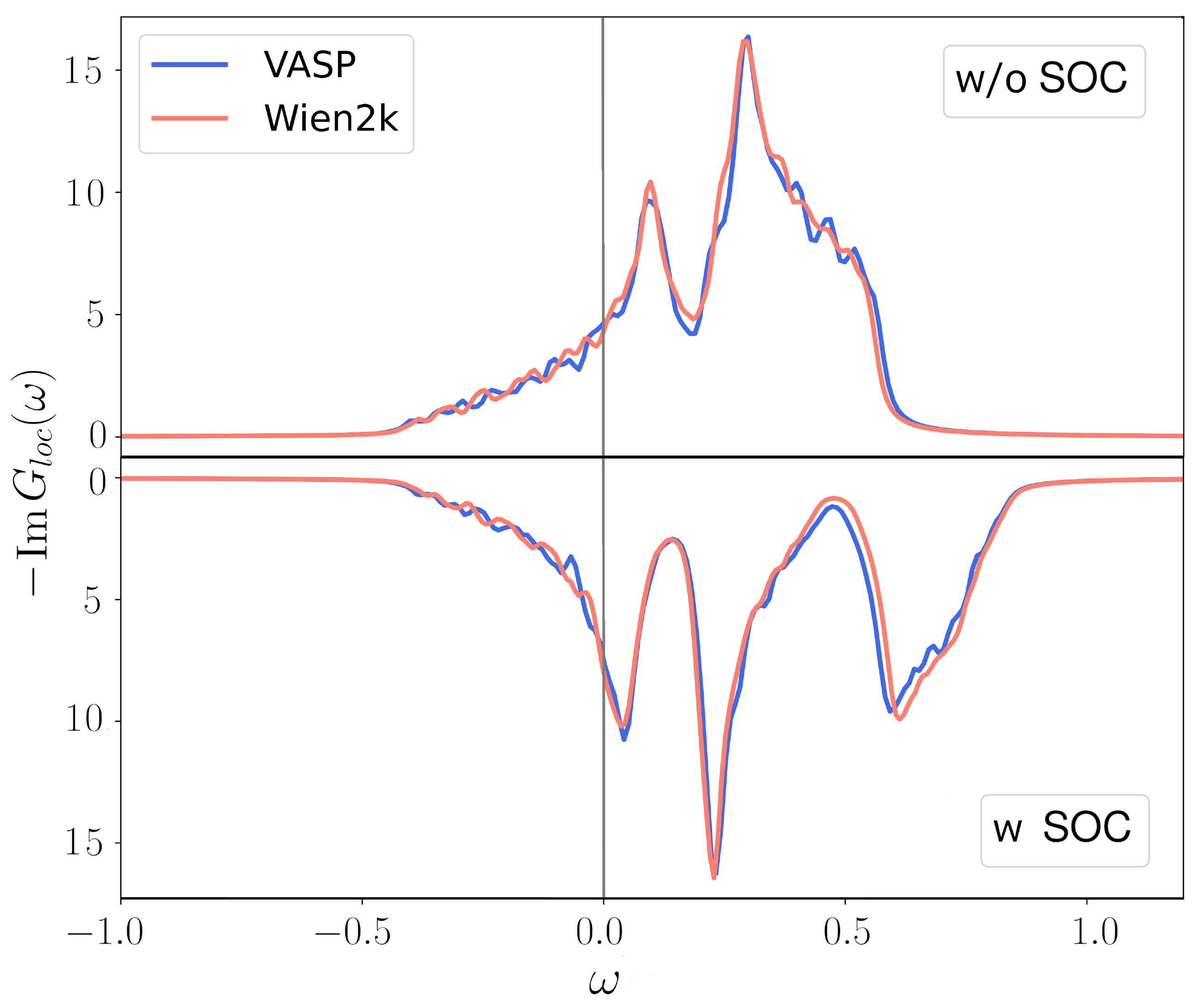}
\caption{Comparison between VASP and Wien2k projected Kohn-Sham spectral function $-\text{Im} G_\text{loc}(\omega)$, equivalent to the DOS, for the calculations without SOC (top panel) and with SOC (bottom panel). 
}
\label{fig:1} 
\end{figure}
The Anderson impurity problem was solved with a state-of-the-art continuous-time quantum Monte-Carlo solver within the hybridization expansion (CT-HYB)~\cite{PhysRevLett.97.076405,PhysRevB.74.155107} as implemented in the TRIQS package~\cite{SETH2016274}. This method is particularly suitable for the present study, as the chosen temperature is well above the critical values where fermionic sign problems become problematic. Every calculation was initialized with $10^5$ warm-up cycles, followed by a set of $\sim 10^7$ measurements. We treated the noisy high-frequency tail of the self energy by a polynomial fit. 
To further reduce the computational cost, the SOC calculations were performed in the numerical $J_\text{eff}$ basis, which is obtained by diagonalizing the local atomic Hamiltonian and restricting the correlated orbitals to the $J_\text{eff}=1/2$ and $3/2$ orbitals, effectively projecting out the e$_g$ states (see Ref.~\cite{Schnait2022}).
The double-counting correction was included in the fully localized limit, and, lastly, an analytic continuation of the imaginary-frequency Green’s functions was performed using the TRIQS/MAXENT code~\cite{PhysRevB.96.155128}.

\begin{figure*}[thp!]
  \begin{center}
  \includegraphics[width=0.9\textwidth,clip=true]{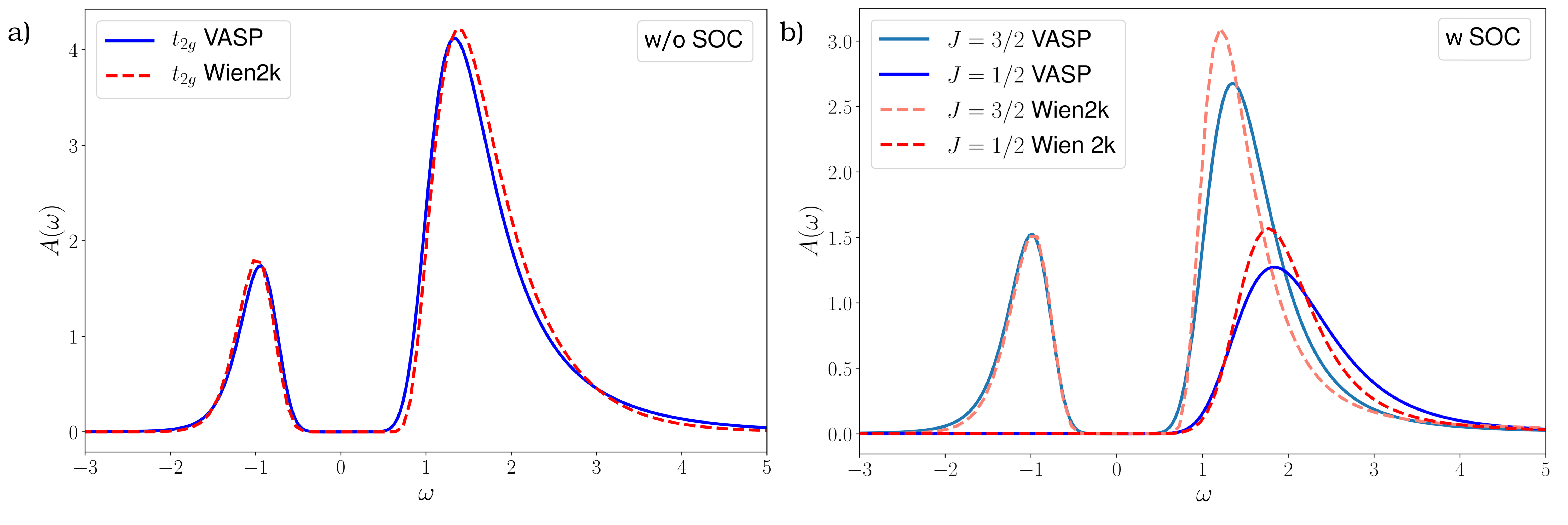}
    \end{center}
\caption{Comparison between the spectral functions of the paramagnetic BNOO between VASP and Wien2k for the case without SOC (a) and with SOC (b). The former shows a single peak structure in the UHB, consequence of the t$_{2g}$ orbital structure, while in the latter there is the appearance of a double-peak structure, whose character is a mixture of $J_\text{eff} = 3/2$ and $J_\text{eff} = 1/2$. The calculations are done with Kanamori interaction values of $U=3.55$\,eV and $J=0.37$\,eV.}
\label{fig:2} 
\end{figure*}

\section{IV. Results}

\subsection{A. Correlated Subspace}
\label{Sec:4a}

In this section, we present our results for the correlated subspace properties and illustrate how the  spinor PAW projectors compare with the corresponding projectors calculated from Wien2k. The comparison was achieved via computation of the local Green's function $G_\text{loc}(\omega)$, the DOS and local impurity Bloch Hamiltonian after convergence of the DFT self-consistent cycle.

We start our comparison from the calculation of the local Green's function, through which we extracted the Kohn-Sham spectral function~-Im$G_\text{loc}(\omega)$. The agreement between VASP and Wien2k results is excellent  both without and with SOC (see Fig.~\ref{fig:1}). Likewise, the total DOS is very well reproduced (see supplementary material). 

Moving to the local Hamiltonian properties, we evaluated the effective atomic levels via diagonalization of the Kohn-Sham Hamiltonian projected onto the local correlated space, as obtained from 
\begin{equation}
H_{mm'}= \sum_{\mathbf{k},\nu\nu'}  {P}_{m \nu} (\mathbf{k})  H_{\nu \nu'} (\mathbf{k})  \big( {P}_{\nu' m'} (\mathbf{k}) \big)^* \ ,
\end{equation}
where $H_{\nu \nu'}(\mathbf{k}) = \epsilon_{\nu} (\mathbf{k}) \delta_{\nu \nu'}$ and $\epsilon_{\nu} (\mathbf{k})$ are the Kohn-Sham eigenvalues. 

The results without SOC provide the expected crystal field levels, i.e. a  sixfold degenerate (including spin multiplicity) t$_{2g}$  ground state multiplet well separated in energy ($\Delta_{CF} \sim~4.7$\,eV) from the excited fourfold degenerate e$_g$ states, with the value of $\Delta_{CF}$ in very good agreement both with Wien2k results and experimental measurements~\cite{Kesavan2020}.
Since the t$_{2g}$ orbitals are degenerate, the occupation of this multiplet is $1/6$ per orbital+spin channel.

Moving to the case with SOC, our calculations of the effective atomic levels indicate the formation of total angular momentum eigenstates, with the $J_\text{eff} = 3/2$ low-energy quadruplet being the ground-state multiplet. The first excited doublet with  $J_\text{eff}  = 1/2$, is separated in energy by $\Delta E \sim~0.44$\,eV. From this value, we extracted the SOC constant  $\lambda \approx 0.3$\,eV, that is consistent with previous theoretical results~\cite{mosca2021interplay}. 
The filling of the ground-state multiplet is close to 0.25 per spin-orbital level, even though a small inter-mixing with $J_\text{eff} = 1/2$ is observed (see Tab.~\ref{Tab:1} for the schematic comparison of VASP and Wien2k energy levels and occupations).

\begin{table}[b]
\caption{\label{Tab:1}Comparison between VASP and Wien2k energy levels and fillings for the SOC correlated subspace. }
\centering
		\begin{ruledtabular}
			\renewcommand{\arraystretch}{1.2}
			\begin{tabular}{l | c c | c c }
				& \multicolumn{2}{c |}{VASP} & \multicolumn{2}{c}{Wien2k} \\ \hline 
				 Levels & Energy (eV) & Filling (e)  & Energy (eV) & Filling (e)\\   
				\hline
				J$_\text{eff}$ = 3/2 & 0.32 & 0.24 & 0.31 & 0.23 \\
				J$_\text{eff}$ = 1/2  & 0.76 & 0.04 & 0.78 & 0.05\\
				e$_g$ & 5.54 & 0.00 & 5.59 & 0.00 \\				
			\end{tabular}
		\end{ruledtabular}
\end{table}

\subsection{B. Metal Insulator Transition}
\label{Sec:4b}

Moving to the electronic properties of BNOO, we find from DFT+DMFT calculations an insulating phase in both the non-SOC and the SOC cases. The estimated band gaps are $\approx 1$\,eV for both, albeit with qualitative differences observed in the respective spectral functions, see Fig.~\ref{fig:2}. 
Indeed, for the non-SOC case, the UHB retains the full t$_{2g}$ character, and shows a single peak feature centered around 1.5\,eV above the Fermi energy. On the contrary, the SOC results exhibit a double-peak structure originating from the combination of the $J_\text{eff} =3/2$ UHB and the $J_\text{eff} = 1/2$ states. Both spectral functions are qualitatively in very good agreement when compared to the Wien2k-based results. The slight differences in the calculation including SOC are likely a consequence of artifacts from the analytic continuation.

\begin{figure*}[th]
  \begin{center}
  \includegraphics[width=1.0\textwidth,clip=true]{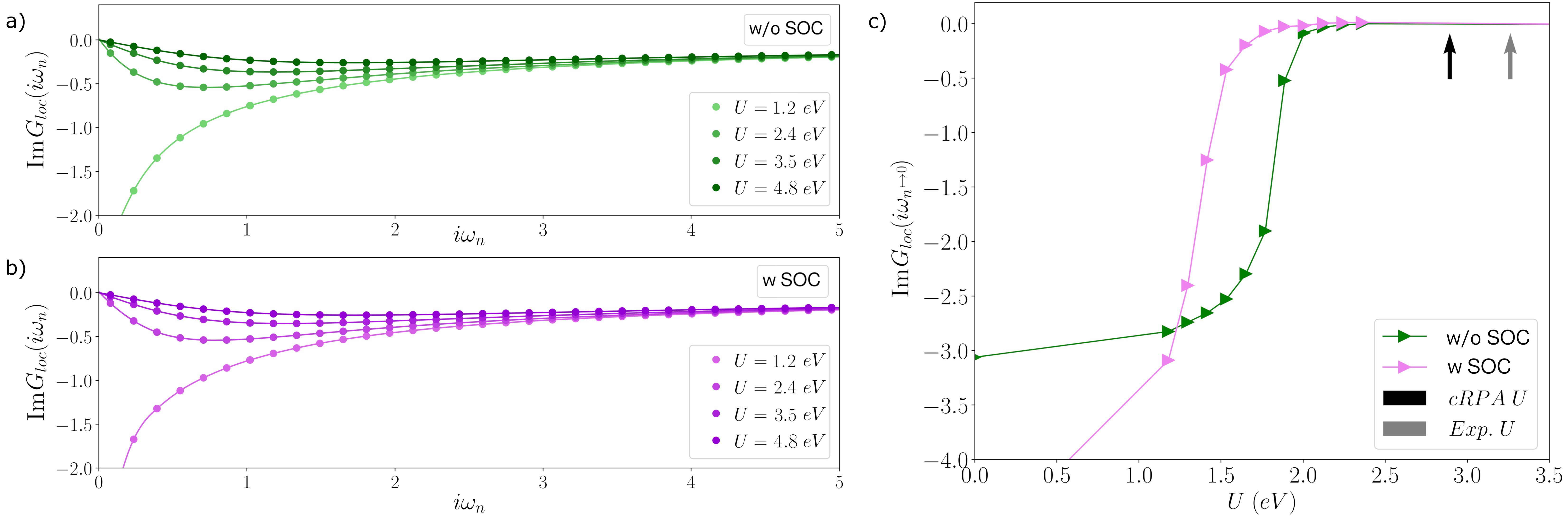}
    \end{center}
\caption{\label{fig:4} The metal-insulator transition as observed in $\text{Im} G_\text{loc}(i\omega_n)$ for values of $U$ between 1.2 and 4.8\,eV, for the cases without SOC (a) and with SOC (b). The transition is apparent in the limit of Matsubara frequencies going to zero. Our refined calculations within the transition region (between 1.0\,eV and 2.5\,eV) (c) highlight a small dependence of the transition on SOC, whose impact on the MIT is to shift the phase boundary to lower interactions by about 0.3\,eV. The phase boundary lies, however, in both cases far below the experimental and computed values of $U$, indicated with arrows.}
\end{figure*}

In order to investigate the nature of BNOO and the influence of SOC, we employed a series of calculations at different values of $U$ and $J$, while keeping the $J/U$ ratio fixed to 0.16.  Our results show, for both non-relativistic and relativistic calculations, a MIT transition between $U=1$\,eV and $U=2$\,eV, see Fig.~\ref{fig:4}. This can be seen in the imaginary part of the interacting local Green's function $\text{Im}\,G_\text{loc}(i\omega_n)$, when taking the limit of $i\omega_n \to 0$. Refined analysis done at several values of $U$ in the proximity of the transition region reveal that the critical interaction without SOC is at about 1.6\,eV, whereas the transition is at roughly $U=1.2$\,eV when SOC is taken into account. This is shown in Fig.~\ref{fig:4} c).
It is within this region of $U=1.2$\,eV to 1.6\,eV that SOC actively affects, and to some extent sustains the MIT, as it opens the gap at lower $U$ values. 

\begin{table}[b]
\caption{\label{Tab:2} Values of the screened Coulomb interactions $U_{ij}$ and $J_{ij}$ for BNOO from cRPA calculation in the "$t_{2g}$/$t_{2g}$" scheme, with an overall value of $U=2.9$\,eV and $J=0.2$\,eV. The calculations were done using the PBE functional. }
	\begin{center}
			\renewcommand{\arraystretch}{1.2}
			\begin{ruledtabular}
			\begin{tabular}{| c | c c c | c c c |}
				 \multicolumn{1}{c}{} & \multicolumn{3}{c}{$U_{ij}$ (eV)} & \multicolumn{3}{c}{$J_{ij}$ (eV)} \\ 
				 \cline{2-7}
				 \multicolumn{1}{c |}{} & d$_{xy}$ & d$_{xz}$ & d$_{yz}$ & d$_{xy}$ & d$_{xz}$ & d$_{yz}$  \\ 
				\hline
				d$_{xy}$ & 2.933  &  2.343 &    2.343 &    &  0.204 & 0.204\\
				d$_{xz}$ & 2.343  &  2.933 &    2.343 & 0.204  &  & 0.204 \\
				d$_{yz}$ & 2.343  &  2.343 &    2.933 & 0.204  &  0.204 & \\
			\end{tabular}
			\end{ruledtabular}			
		\end{center}
\end{table}

Given that the properties of BNOO are closely linked to the strength of its onsite Coulomb interaction, we decided to compute the screened interaction parameters, $U$ and $J$, via cRPA. The calculation was conducted using the "$t_{2g}/t_{2g}$" scheme~\cite{PhysRevB.86.165105} based on a non-magnetic and non-relativistic band structures, with the correlated subspace subjected to wannierization, performed within a window $\mathcal{W} = [-1,1]$\,eV centered around the Fermi level. The results are shown in Tab.\ref{Tab:2}. 
\textcolor{black}{Our cRPA~results lead to a on-site averaged intra-orbital Coulomb interaction $U=2.9$\,eV and an averaged Hund's exchange $J=0.20$\,eV. These values are slightly lower than the experimentally calculated $U=3.3$\,eV~\cite{Erickson2007} or from other theoretical works~\cite{mosca2021interplay}. However, all these estimates lead to sizeable Coulomb interaction parameters, and this indicates that electron correlation plays a crucial role, exerting a greater influence than expected on 5d transition metals~\cite{Liu2015}}. 

{\color{black} We suggest that the strong electronic correlations to be mostly driven by the small ratio between bandwidth $W \sim 1$\,eV and on site Coulomb term $U\sim 3$\,eV, brought by the relatively large inter-atomic distance provided by the double perovskite structure. This ratio $W/U$ is much smaller than in ordinary perovskites, pushing the system easily into the insulating regime, both for calculations with and without SOC. As discussed in figures \ref{fig:2} and~\ref{fig:4}, SOC helps this transition but is not detrimental. Besides, other factors can also have an impact on the strength of the electronic correlation, namely hybridization effects~\cite{Gangopadhyay2015, mosca2021interplay}.   
As has been exemplified in the 5d compound Sr$_2$IrO$_4$, Coulomb interactions are larger in this 5d compound as compared to the 4d isovalent and isostructural Sr$_2$RhO$_4$~\cite{Martins_2017}. However, this effect on the magnitude of $U$ is small compared to the effect of the already small $W/U$ ratio.} 
Although the cRPA results serve as only an estimate of the actual interaction values, it is evident that the genuine interaction values for BNOO are much higher than the range where SOC significantly influences the MIT (Fig.~\ref{fig:4}). As BNOO's insulating phase is present regardless of relativistic effects, it should be classified as a more standard Mott insulator.

Furthermore, when exploring different potential explanations for the insulating behavior in the paramagnetic phase, such as disorder and symmetry breaking, electron correlation continues to emerge as a critical factor. Recent works on transition metal oxides have highlighted that a polymorphous description of DFT can lift the degeneracy of the d orbitals and allow for the formation of a band gap in 3d binary oxides~\cite{PhysRevB.102.045112,PhysRevB.97.035107,PhysRevMaterials.5.104410}. Our results with a spin-paramagnetic configuration show that even within a polymorphous approach to DFT this insulating phase is not observed without any correction in the form of Hubbard U (See Section B of the Supplemental Materials).

Our results shed light on the insulating nature of BNOO and highlight possible future pathways. Understanding the role played by magnetic interactions in combination with correlation effects, possible local distortions and SOC effect is an ongoing research topic, and further studies are needed. The present implementation of the  spinor DMFT will allow to further clarify the complex interplay of these phenomena and address the phase transition of the canted AFM phase.

\section{V. Conclusion}
\label{sec:5}

To conclude, we have extended the TRIQS/DFTTools interface with VASP to account for  spinor and spin-orbit coupled DFT+DMFT calculations, and tested this implementation in the study of the MIT in the 5d$^1$ double perovskite BNOO. Our calculations in the paramagnetic region at room temperature prove that the SOC is not necessary for opening the gap in this compound. Its insulating phase originates from the action of correlation effects, making this material a Mott insulator. We have obtained qualitative and quantitative differences between the non-SOC and SOC calculations, whose trustworthiness could be proven by comparison with experimental measurements. Lastly, our implementation of  spinor projectors within the DMFT framework opens up further possibilities to study these strong SOC compounds from a magnetic perspective, including non-collinear magnetic orderings and canted magnetic structures.

\section{Data availability}

\textcolor{black}{The implementation of our current approach can be accessed in the most recent GitHub repository of TRIQS/dft\_tools. Data will be made available on request.}

\section{Acknowledgments}

We acknowledge funding from the Austrian Science Fund (FWF), projects J4698 and Y746. D. Fiore Mosca and L. Celiberti acknowledge the Vienna Doctoral School of Physics. The computational results presented have been achieved using the Vienna Scientific Cluster (VSC). D.Fiore Mosca would like to thank Oleg Peil and Alexander Hampel for the useful discussions.

\bibliography{biblio}

\end{document}


\title{Supplementary material for 
'The Mott transition in the 5d$^1$ compound Ba$_2$NaOsO$_6:$ a DFT+DMFT study with PAW non-collinear projectors'}

\author{Dario Fiore Mosca}
\affiliation{Centre de Physique Th\'eorique, Ecole Polytechnique, CNRS, Institut Polytechnique de Paris,
  91128 Palaiseau Cedex, France}
\affiliation{Coll\`ege de France, 11 place Marcelin Berthelot, 75005 Paris, France}

\author{Hermann Schnait}
\affiliation{Institute of Theoretical and Computational Physics, Graz University ofTechnology, NAWI Graz, 8010 Graz, Austria}

\author{Lorenzo Celiberti}
\affiliation{University of Vienna, Faculty of Physics and Center for Computational Materials Science, Vienna, Austria}

\author{Markus Aichhorn}
\affiliation{Institute of Theoretical and Computational Physics, Graz University ofTechnology, NAWI Graz, 8010 Graz, Austria}

\author{Cesare Franchini}
\affiliation{University of Vienna, Faculty of Physics and Center for Computational Materials Science, Vienna, Austria}
\affiliation{Department of Physics and Astronomy "Augusto Righi", Alma Mater Studiorum - Universit\`a di Bologna, Bologna, 40127 Italy}

\date{\today}	
\maketitle

\section{Density of States}

The following figure shows the comparison between the total DFT density of states, and the one obtained from the PAW non-collinear projectors from TRIQS.

\begin{figure}[h]
    \centering
    \includegraphics[width=0.6\linewidth]{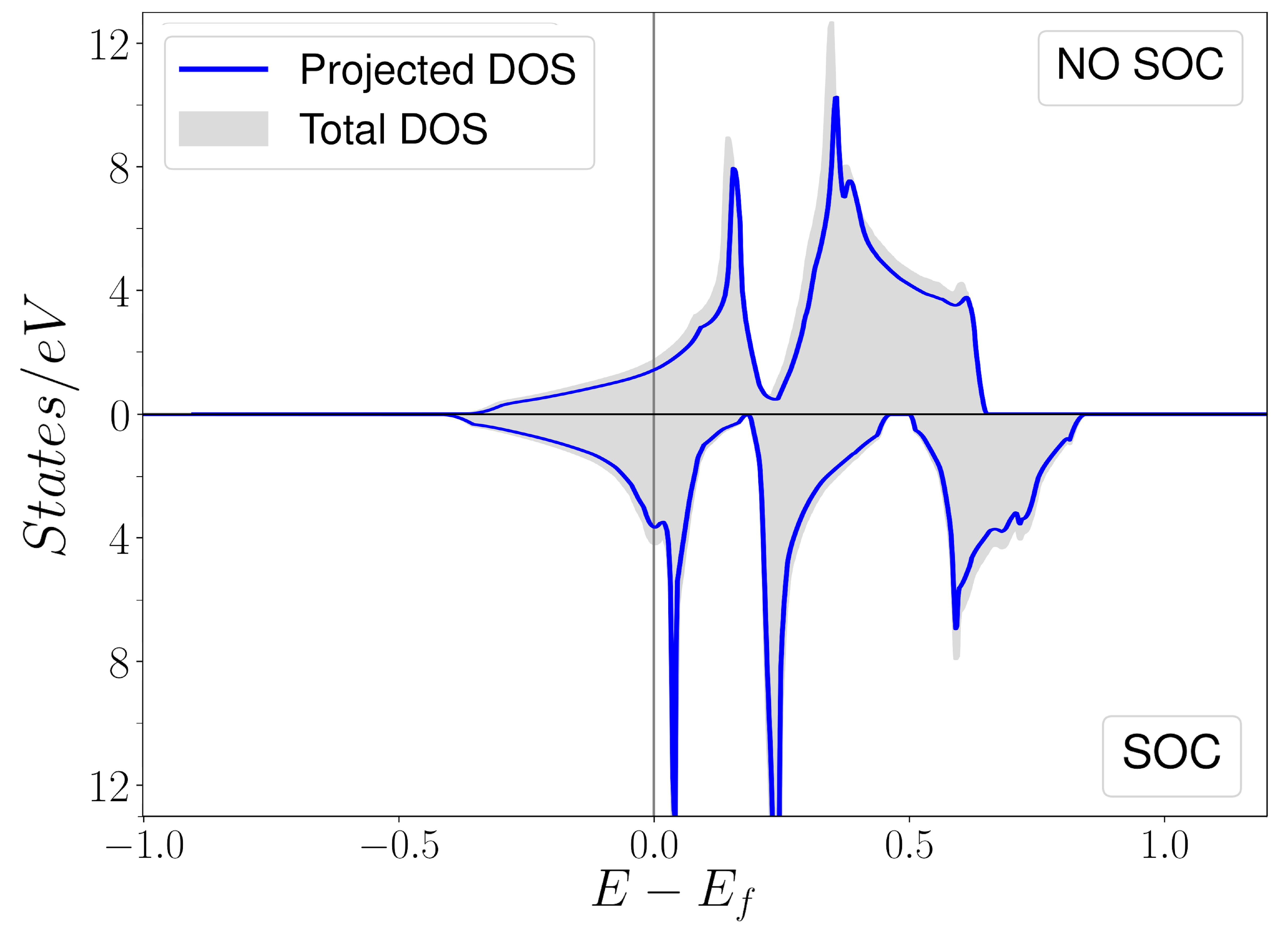}
    \caption{Comparison of the DOS from DFT and from PAW projectors.}
    \label{fig:my_label}
\end{figure}

\clearpage

\section{DFT Computational Details}

This section provides a summary of the computational methodology used for calculating the metallic and insulating density of states discussed in the main text. To perform the calculations, we utilized the VASP program and the Perdew-Burke-Ernzerhof approximation of the exchange correlation functional, with an 8 f.u. supercell of dimensions  $\sqrt{2}a\times \sqrt{2} \times a$, where $a$ is equal to 8.287 \AA. We incorporated Dudarev's DFT+U scheme, applying a $U_{eff} = U-J = 3.4$,eV, consistent with previous research~\onlinecite{mosca2021interplay}. We further deactivated all symmetries and activated the non-collinear routine, while using an energy cutoff of 600\,eV with a k-mesh of $6\times6\times4$. 
Non-collinear magnetic orderings were implemented through a penalty energy functional as described in a related work of Dudarev~\onlinecite{dudarev_parametrization_2019} with a value of the penalty energy constant $\lambda$ = 10, which allowed calculations of precision of $10^{-4}$ eV. The "canted Antiferromagnetic Configuration" referred to in the main text was initiated following the guidelines of Ref.~\onlinecite{mosca2021interplay}. 

\textcolor{black}{In case of the Jahn-Teller distorted structure, we allowed all atoms to relax to their optimal positions. For this purpose we used a a quasi-Newton algorithm with a step width of 0.5 \AA~ and an energy convergence of 10$^{-2}$ eV. The obtained structure in cif format is reported at the end of the supplementary materials. It is important to clarify that, although JT distortions are present as a consequence of the orbital degeneracy and electron occupancy of the $t_{2g}$ states~\cite{kugel1982jahn}, the solution derived under these conditions is not physically representative.  The omission of SOC effect prevents the spin-orbital entanglement, which is critical for the shift of the system from being characterized by real cubic harmonics to spherical complex ones. Particularly in the context of a 5d$^1$ configuration,  the inclusion of SOC is expected to reduce the amplitude of the JT modes~\cite{PhysRevX.10.031043, PhysRevB.105.205142}.}

\subsection{DFT + $U$ + AFM phase}

To provide a comprehensive overview, we present an additional insulating solution that does not involve the spin-orbit coupling effect. The sole distinction from the previous calculations is the adoption of a type-I antiferromagnetic phase along the [001] direction with magnetic moments oriented in the $z$-direction and without structural distortions.

\begin{figure}[h]
\begin{center}
    \includegraphics[width=0.6\linewidth]{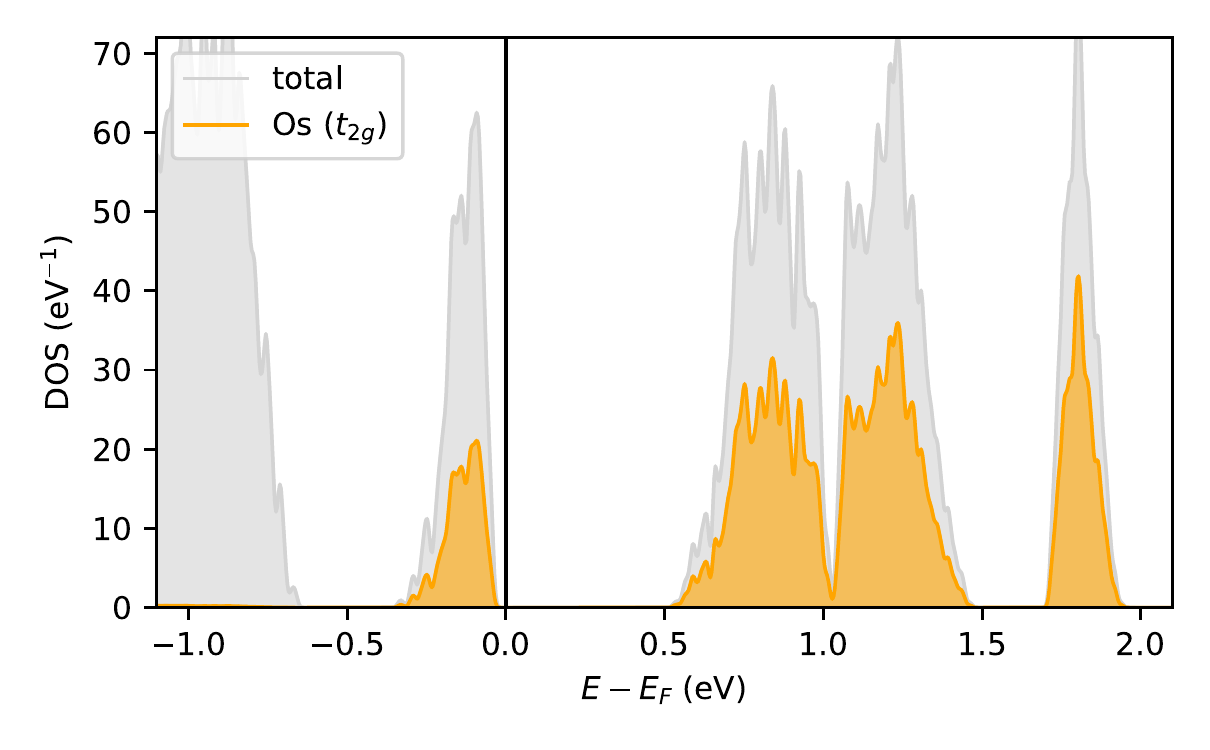}
    \caption{Density of states of the canted AFM configuration with DFT+$U$ at $U= 3.4$\,eV.}
\end{center}
\end{figure}

\subsection{DFT polymorphous representation using a paramagnetic supercell}

To further assess the role of electronic correlations in this compound, we have explored the use of a supercell approach that has recently been employed for obtaining paramagnetic and insulating solutions in compounds that would be otherwise metallic within the standard DFT framework~\cite{PhysRevB.100.035119, PhysRevB.102.045112}.

\begin{figure}
    \centering
    \includegraphics[width=0.4\linewidth]{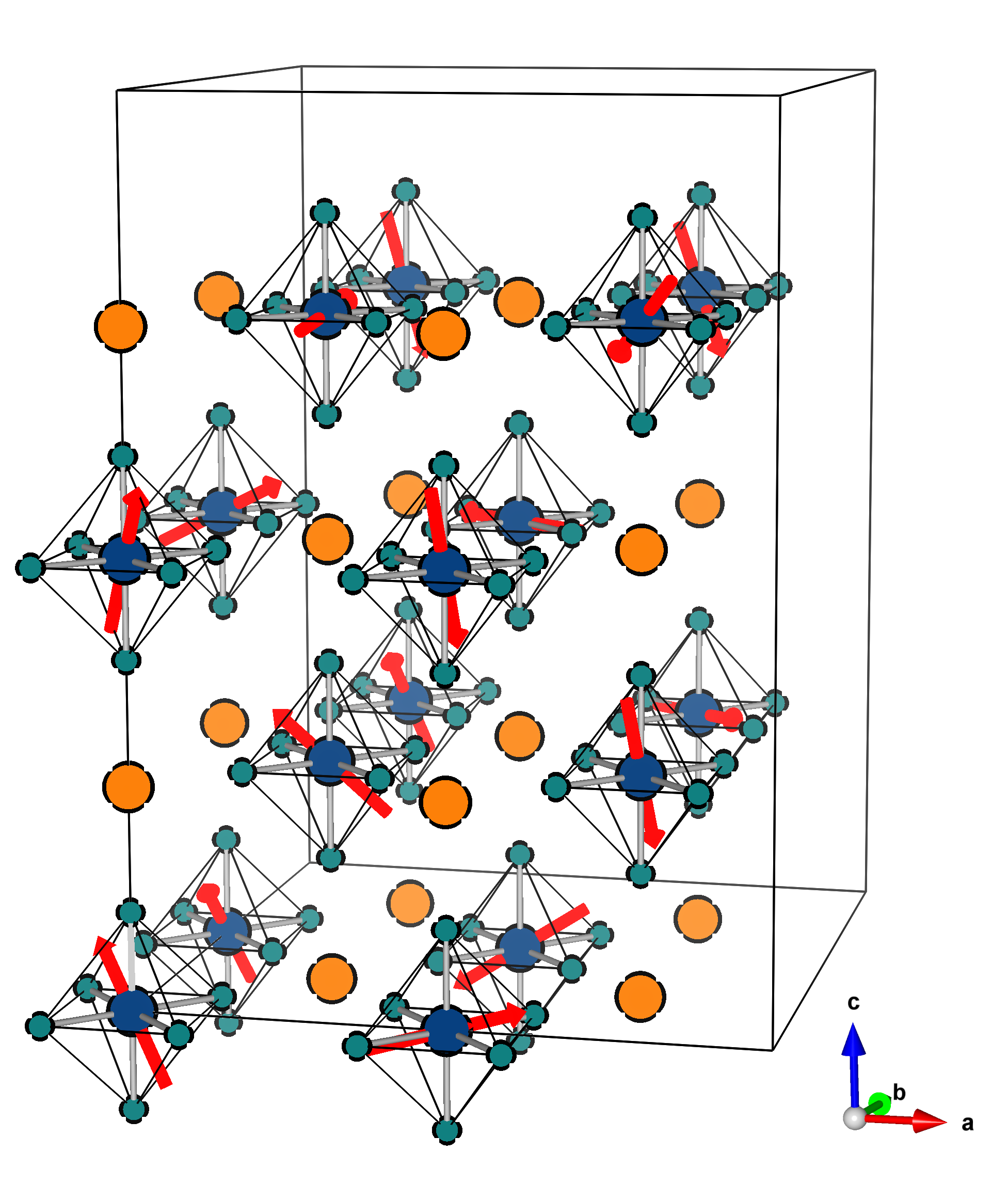}
    \caption{Supercell containing 16 f.u. with magnetic moments initialised randomly used for the DFT without and with SOC (and without U). The magnetic moments on the in-equivalent osmium sites are also thereby shown.}
    \label{fig:paramagnetic}
\end{figure}

\noindent We have computed the Density of States for both DFT and DFT+SOC calculations using a 16 f.u. super cell with lattice constants $  a\sqrt{2} \times a \sqrt{2} \times 2a$, consisting of 160 atoms. This cell is shown in Figure~\ref{fig:paramagnetic}. 
The paramagnetic phase was here simulated with both PBE as well as SCAN potentials like in previous theoretical works~\cite{PhysRevB.102.045112, PhysRevMaterials.5.104410}. We applied the following procedure: 

\begin{enumerate}
    \item The overall shape of the supercell is kept fixed to the microscopically observed lattice symmetry Fm-3m. 
    \item A spin-paramagnetic configuration was simulated by initializing randomly the direction of the magnetic moments, such that the overall sum in the unit cell is zero. We preferred this implementation with respect to a spin-up spin-down only configuration, as our calculations were performed in a non-collinear setup~\cite{PhysRevB.92.054428}. 

    In our case 14 magnetic moments were randomly initialized, and the remaining 2 were adapted in order to recover the net zero overall magnetic moment of the supercell. A constrained magnetic moment approach on the direction of the moments was also enforced (see Figure~\ref{fig:paramagnetic} for the visualization of the magnetic moments). 
    \item The atoms were allowed to relax to their optimal configuration, while retaining the overall cubic symmetry of the cell. We further tested the case of a fully relaxed unit cell, without finding significant differences.

    A conjugate gradient algorithm was employed for this purposes, with an accuracy on the relaxation of 10$^{-3}$. 
    \item The occupation of the degenerate occupied orbitals was not forced to be the same, also by means of switching all symmetries off. 
    In such manner, there is no symmetry constraint applied for the DFT calculations. We further highlight that, especially in presence of SOC effect, the random initialization of magnetic moments produces a nudge of the atomic displacements and orbital occupations, as a consequence of the coupling between the spin and orbital momenta. 
\end{enumerate}
A k-mesh of $2\times2\times1$ was employed for the structural minimization, while a k-mesh of $3\times3\times2$ for the Density of States calculation. An energy cutoff on the plane wave expansion of 600 eV was adopted, together with a convergence criterion for the energy of 10$^{-5}$ eV.  We have not tested larger supercell structures due to the computational  cost associated with this type of calculation. 

\begin{figure}
    \centering
    \includegraphics[width=0.7\linewidth]{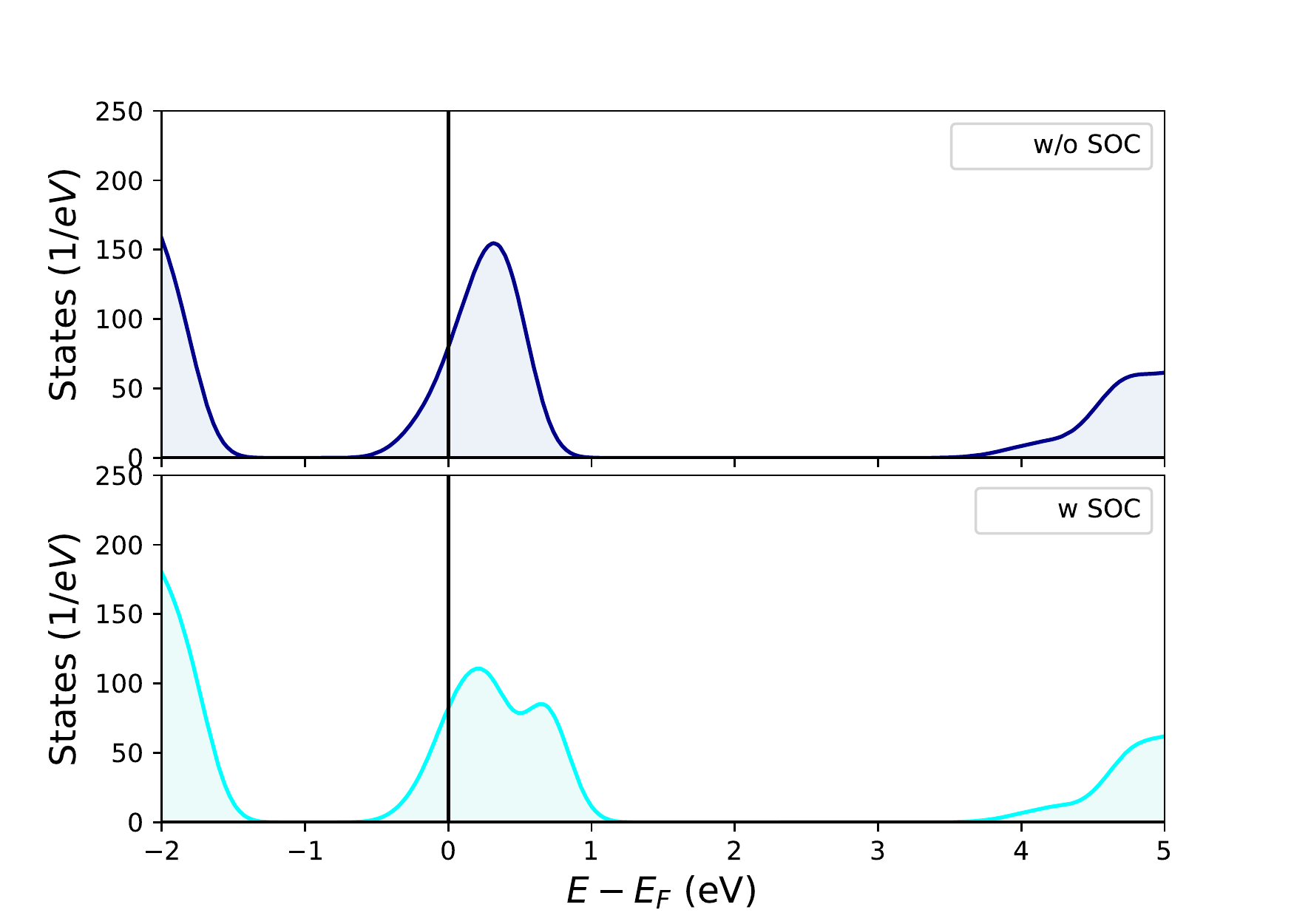}
    \caption{Density of States for the DFT paramagnetic calculation without and with SOC effect with SCAN potential. Both DOS provide a metallic solution.}
    \label{fig:scan_supercell}
\end{figure}

\begin{figure}
    \centering
    \includegraphics[width=0.7\linewidth]{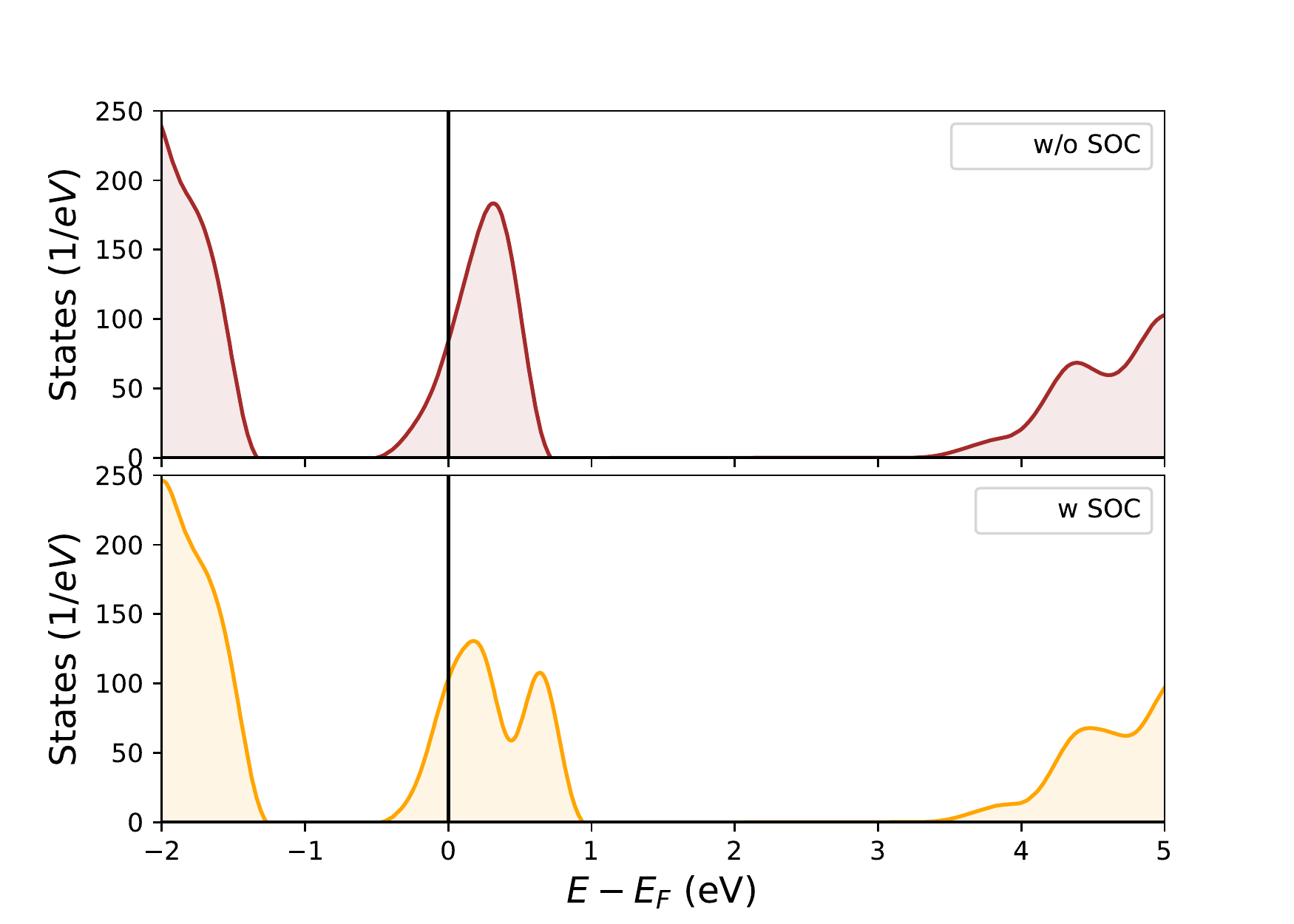}
    \caption{Density of States for the DFT paramagnetic calculation without and with SOC effect using PBE potential. Both DOS provide a metallic solution.}
    \label{fig:supercell}
\end{figure}

\noindent Our results show that in absence of +U corrections, the DOS is always metallic, both with and without SOC, as well as for PBE and SCAN potentials (see figures ~\ref{fig:scan_supercell} and ~\ref{fig:supercell}). The onsite Coulomb interaction seems to be critical for the occurrence of the paramagnetic insulating phase in this compound.  

\clearpage
\section{VASP DMFT computational details}

VASP employs a universal reference frame for the spin and orbital components. In this case, the spin quantization axis was aligned with the global z-axis direction. As a result, we had to implement a rotation to align the local reference frame of the Os-O octahedra with the global reference frame. To achieve this alignment, we derived the transformation using the diagonalization of the effective atomic levels. 
\footnotesize
\begin{equation}
\begin{pmatrix}
 0 & 0 & 0 & -0.817 & 0 &  0 &  0 & -0.577 &  0 &  0 \\
 0 & 0 & 0.817 &  0 & 0 &  0 &  0 &  0 &  0 &  0.577 \\
 0 & 0 & 0 &  0 & 0.577 &  0 &  0.817 &  0 &  0 &  0 \\
 0 & 0 & 0 &  0 & 0 & -0.577 &  0 &  0 & -0.817 &  0 \\
 1.000 & 0 & 0 &  0 & 0 &  0 &  0 &  0 &  0 &  0 \\
 0 & 1.000 & 0 &  0 & 0 &  0 &  0 &  0 &  0 &  0 \\
 0 & 0 & 0 & -0.577 & 0 &  0 &  0 &  0.817 &  0 &  0 \\
 0 & 0 & 0.577 &  0 & 0 &  0 &  0 &  0 &  0 & -0.817 \\
 0 & 0 & 0 &  0 & 0.817 &  0 & -0.577 &  0 &  0 &  0 \\
 0 & 0 & 0 &  0 & 0 & -0.817 &  0 &  0 &  0.577 &  0 \\
\end{pmatrix}
\end{equation}
\normalsize
which leads to the following order of the orbitals: 
$\big( d_{xy,\uparrow},d_{xy,\downarrow},d_{yz,\uparrow},d_{yz,\downarrow},d_{z^{2},\uparrow},d_{z^{2},\downarrow},d_{xz,\uparrow},d_{xz,\downarrow},d_{x^{2}-y^{2},\uparrow},d_{x^{2}-y^{2},\downarrow} \big)$. 
In TRIQS/operators the construction of the four-index U matrix follows the conventional order: $\big( d_{xy,\uparrow},d_{yz,\uparrow},d_{z^{2},\uparrow},d_{xz,\uparrow},d_{x^{2}y^{2},\uparrow},d_{xy,\downarrow},d_{yz,\downarrow},d_{z^{2},\downarrow},d_{xz,\downarrow},d_{x^{2}-y^{2},\downarrow} \big)$. A mapping between the two basis sets was adopted to make the interacting Hamiltonian of the Anderson Impurity problem consistent.

\section{Wien2k DFT computational details}

To validate the results of our non-collinear projectors from VASP, we performed a similar DFT(+SOC) calculation using the LAPW basis set as implemented in Wien2k.\cite{Blaha_2020_PAPER}
In contrast to VASP, Wien2k allows the inclusion of SOC in collinear calculations by using a variational approach.
Wien2k has been used for DFT+DMFT calculations in the past, both using maximally localized Wannier function as well as projective Wannier functions.
For comparison with our non-collinear projectors from VASP, we used projective Wannier functions as implemented in dmftproj in TRIQS/DFTtools.\cite{Aichhorn2016} 
In the non-relativistic calculation, the projection was performed onto the Os t$_{2g}$ manifold, utilizing the symmetry operations employed by Wien2k.
In the relativistic case, e$_g$ and t$_{2g}$ are no longer irreducible representations of the $d$-shell, thus the projection was performed onto the full $d$ manifold.

The Wien2k DFT calculations have been performed using $5000$ k-points in the irreducible Brillouin zone using the PBE functional. After convergence was reached, the states within a energy window of $[-1, 5.8]~\text{eV}$ ($[-0.8, 6.8]~\text{eV}$ for the relativistic case) around the Fermi level were projected onto the t$_{2g}$ ($d$) shell of the Os atom.

\section{JT-distorted Structure}
\footnotesize

\begin{verbatim}
    
    

#======================================================================
# CRYSTAL DATA
#----------------------------------------------------------------------
data_VESTA_phase_1

_chemical_name_common                  'BNOO                                  '
_cell_length_a                         11.719587
_cell_length_b                         11.719587
_cell_length_c                         8.287000
_cell_angle_alpha                      90.000000
_cell_angle_beta                       90.000000
_cell_angle_gamma                      90.000000
_cell_volume                           1138.208861
_space_group_name_H-M_alt              'P 1'
_space_group_IT_number                 1

loop_
_space_group_symop_operation_xyz
   'x, y, z'

loop_
   _atom_site_label
   _atom_site_occupancy
   _atom_site_fract_x
   _atom_site_fract_y
   _atom_site_fract_z
   _atom_site_adp_type
   _atom_site_U_iso_or_equiv
   _atom_site_type_symbol
   Na1        1.0     0.000000     0.500000     0.500000    Uiso  ? Na
   Na2        1.0     0.500000     0.500000     0.500000    Uiso  ? Na
   Na3        1.0     0.500000     0.000000     0.500000    Uiso  ? Na
   Na4        1.0     0.750000     0.750000     0.000000    Uiso  ? Na
   Na5        1.0     0.750000     0.250000     0.000000    Uiso  ? Na
   Na6        1.0     0.000000     0.000000     0.500000    Uiso  ? Na
   Na7        1.0     0.250000     0.250000     0.000000    Uiso  ? Na
   Na8        1.0     0.250000     0.750000     0.000000    Uiso  ? Na
   Os1        1.0     0.000000     0.000000     0.000000    Uiso  ? Os
   Os2        1.0     0.500000     0.000000     0.000000    Uiso  ? Os
   Os3        1.0     0.500000     0.500000     0.000000    Uiso  ? Os
   Os4        1.0     0.750000     0.250000     0.500000    Uiso  ? Os
   Os5        1.0     0.750000     0.750000     0.500000    Uiso  ? Os
   Os6        1.0     0.000000     0.500000     0.000000    Uiso  ? Os
   Os7        1.0     0.250000     0.750000     0.500000    Uiso  ? Os
   Os8        1.0     0.250000     0.250000     0.500000    Uiso  ? Os
   Ba1        1.0     0.000000     0.250000     0.750000    Uiso  ? Ba
   Ba2        1.0     0.500000     0.250000     0.750000    Uiso  ? Ba
   Ba3        1.0     0.500000     0.750000     0.750000    Uiso  ? Ba
   Ba4        1.0     0.750000     0.500000     0.250000    Uiso  ? Ba
   Ba5        1.0     0.750000     0.000000     0.250000    Uiso  ? Ba
   Ba6        1.0     0.000000     0.750000     0.750000    Uiso  ? Ba
   Ba7        1.0     0.250000     0.000000     0.250000    Uiso  ? Ba
   Ba8        1.0     0.250000     0.500000     0.250000    Uiso  ? Ba
   Ba9        1.0     0.000000     0.750000     0.250000    Uiso  ? Ba
   Ba10       1.0     0.500000     0.750000     0.250000    Uiso  ? Ba
   Ba11       1.0     0.500000     0.250000     0.250000    Uiso  ? Ba
   Ba12       1.0     0.750000     0.000000     0.750000    Uiso  ? Ba
   Ba13       1.0     0.750000     0.500000     0.750000    Uiso  ? Ba
   Ba14       1.0     0.000000     0.250000     0.250000    Uiso  ? Ba
   Ba15       1.0     0.250000     0.500000     0.750000    Uiso  ? Ba
   Ba16       1.0     0.250000     0.000000     0.750000    Uiso  ? Ba
   O1         1.0     0.862463     0.363592     0.500435    Uiso  ? O
   O2         1.0     0.364074     0.366082     0.499509    Uiso  ? O
   O3         1.0     0.362344     0.864628     0.500641    Uiso  ? O
   O4         1.0     0.615139     0.615254    -0.000986    Uiso  ? O
   O5         1.0     0.612465     0.114801     0.999247    Uiso  ? O
   O6         1.0     0.865633     0.864649     0.499233    Uiso  ? O
   O7         1.0     0.114833     0.115444     0.000289    Uiso  ? O
   O8         1.0     0.112369     0.613739     1.000283    Uiso  ? O
   O9         1.0     0.137398     0.635498     0.499586    Uiso  ? O
   O10        1.0     0.634649     0.635193     0.500958    Uiso  ? O
   O11        1.0     0.637695     0.136352     0.499242    Uiso  ? O
   O12        1.0     0.884890     0.884793     0.999805    Uiso  ? O
   O13        1.0     0.887509     0.386214     0.999933    Uiso  ? O
   O14        1.0     0.135695     0.134000     0.500057    Uiso  ? O
   O15        1.0     0.385302     0.384383     1.000576    Uiso  ? O
   O16        1.0     0.387835     0.885087     0.000450    Uiso  ? O
   O17        1.0     0.134729     0.363365     0.501657    Uiso  ? O
   O18        1.0     0.636574     0.365338     0.499360    Uiso  ? O
   O19        1.0     0.635664     0.864791     0.501321    Uiso  ? O
   O20        1.0     0.886863     0.615576     0.999249    Uiso  ? O
   O21        1.0     0.885819     0.114033     0.001684    Uiso  ? O
   O22        1.0     0.137115     0.864938     0.499564    Uiso  ? O
   O23        1.0     0.386451     0.114591    -0.000263    Uiso  ? O
   O24        1.0     0.385738     0.613600     0.001800    Uiso  ? O
   O25        1.0     0.864090     0.635211     0.498257    Uiso  ? O
   O26        1.0     0.363054     0.635232     0.500405    Uiso  ? O
   O27        1.0     0.365530     0.136665     0.498932    Uiso  ? O
   O28        1.0     0.613340     0.885199     1.000378    Uiso  ? O
   O29        1.0     0.614159     0.386414     0.998278    Uiso  ? O
   O30        1.0     0.863199     0.134419     0.500748    Uiso  ? O
   O31        1.0     0.113559     0.384815     0.000728    Uiso  ? O
   O32        1.0     0.114365     0.885972     0.998383    Uiso  ? O
   O33        1.0     0.999800     0.499344     0.769849    Uiso  ? O
   O34        1.0     0.500616     0.498966     0.773016    Uiso  ? O
   O35        1.0     0.501490     0.000246     0.770145    Uiso  ? O
   O36        1.0     0.750347     0.749191     0.274021    Uiso  ? O
   O37        1.0     0.750104     0.250340     0.269686    Uiso  ? O
   O38        1.0     0.001625     0.998978     0.773430    Uiso  ? O
   O39        1.0     0.251094     0.248512     0.273970    Uiso  ? O
   O40        1.0     0.250814     0.749734     0.269732    Uiso  ? O
   O41        1.0     0.999645     0.499699     0.230182    Uiso  ? O
   O42        1.0     0.498909     0.501705     0.226946    Uiso  ? O
   O43        1.0     0.498722     0.000677     0.229843    Uiso  ? O
   O44        1.0     0.749310     0.751557     0.725862    Uiso  ? O
   O45        1.0     0.750283     0.250448     0.730243    Uiso  ? O
   O46        1.0     0.998770     1.000394     0.226588    Uiso  ? O
   O47        1.0     0.249200     0.250740     0.726130    Uiso  ? O
   O48        1.0     0.249005     0.749251     0.730328    Uiso  ? O
\end{verbatim}

\bibliography{biblio}